\title{A SUBLUMINOUS SCHR\"ODINGER EQUATION}
\author{Philip Rosenau and Zeev Schuss\thanks{Department of Applied Mathematics,
Tel-Aviv University, Tel-Aviv 69978,  Israel, e-mail:
 rosenau@post.tau.ac.il,schuss@post.tau.ac.il}}
\documentclass[12pt]{article}
\usepackage{float}
\usepackage{amsmath}
\usepackage{epsfig, graphicx} 

\newcommand{\p}{\partial}

\newcommand{\eps}{\varepsilon}

\newcommand{\ds}{\displaystyle}
\newcommand{\beq}{\begin{eqnarray}}
\newcommand{\beqq}{\begin{eqnarray*}}
\newcommand{\eeq}{\end{eqnarray}}
\newcommand{\eeqq}{\end{eqnarray*}}

\font\bb=msbm10 at 12pt  

\def\rR{\hbox{\bb R}}

\begin{document}
\numberwithin{equation}{section}
\maketitle
\newpage

\begin{abstract}
\vspace*{0.3cm}

The standard derivation of Schr\"odinger's equation from a Lorentz-invariant
Feynman path integral consists in taking first the limit of infinite speed of
light and then the limit of short time slice. In this order of limits the light
cone of the path integral disappears, giving rise to an instantaneous spread of
the wave function to the entire space. We ascribe the failure of
Schr\"odinger's equation to retain the light cone of the path integral to the
very nature of the limiting process: it is a regular expansion of a singular
approximation problem, because the boundary conditions of the path integral on
the light cone are lost in this limit. We propose a distinguished limit, which
produces an intermediate model between non-relativistic and relativistic
quantum mechanics: it produces Schr\"odinger's equation and preserves the zero
boundary conditions on and outside the original light cone of the path
integral. These boundary conditions relieve the Schr\"odinger equation of
several annoying, seemingly unrelated unphysical artifacts, including
non-analytic wave functions, spontaneous appearance of discontinuities,
non-existence of moments when the initial wave function has a jump
discontinuity (e.g., a collapsed wave function after a measurement), the EPR
paradox, and so on. The practical implications of the present formulation are
yet to be seen.
\end{abstract}


\noindent PACS numbers 03.65.-w, 03.65.Ta

\section{Introduction}

The original derivation of Schr\"odinger's equation from Feynman's path
integral postulates a classical non-Lorentz-invariant action \cite{Feynman},
\cite{Schulman}, \cite{Kleinert}, \cite{Keller}. The Lorentz-invariant
(discrete) Feynman path integral \cite{Miura}, which propagates within the
light cone emanating from the support of the initial wave function, is usually
approximated by expanding the Lorentz-invariant action in powers of $\dot x/c$
first ($c$ is the speed of light) and then expanding the path integral in
powers of the time slice $\Delta t$. In this approximation the light-cone
disappears and Schr\"odinger's equation in the entire space is recovered, as in
the non-relativistic case (see details below). If the order of limits is
reversed, the discrete path integral converges to the initial wave function and
does not propagate. This anomaly in the approximation of the path integral can
be ascribed to the observation that the aforementioned power series is a
regular expansion of a two-parameter singular perturbation problem. Therefore,
the construction of a uniform asymptotic approximation to the Lorentz-invariant
Feynman path integral calls for a singular perturbation approach.

In this paper we construct a uniform asymptotic approximation to the
Lorentz-invariant Feynman path integral by identifying a distinguished limit of
large $c$ and small time slice $\Delta t$, such that $c\Delta t\to0$, but
$c\sqrt{\Delta t}\to\infty$. More specifically, as the speed of light $c$ is
constant, and explicitly present in the problem, the time slice $\Delta t$
cannot be assumed arbitrarily small, as explained above. In fact, the
distinguished limit requires that $\Delta t$ be bounded below by $\Delta
t\gg\ds\frac{\hbar}{mc^2}\approx6.2\times10^{-22}$ sec (for an electron). In
this distinguished limit the leading order approximation to the path integral
is a solution of the initial value problem for Schr\"odinger's equation with
zero boundary conditions on and outside the light-cone emanating from the
support of the initial wave function (see Figure \ref{f:Light-cone}). This
subluminous Schr\"odinger equation can be considered an intermediate model
between non-relativistic and relativistic quantum mechanics and it is valid for
the above mentioned time resolution. We stress, however, that it is not our aim
to construct a comprehensive relativistic quantum mechanics theory (in, say,
the sense of \cite{Berestetskii}).

Notably, the emergence of the light cone relieves Schr\"odinger's equation of
several annoying, seemingly unrelated mathematical and unphysical artifacts.
These include non-analytic wave functions, spontaneous appearance of
discontinuities \cite{Peres}, non-existence of moments when the initial wave
function has a jump discontinuity (e.g., a collapsed wave function after a
measurement) \cite{unidirect}, the EPR paradox \cite{EPR} (relativistic quantum
mechanics theory not withstanding), and so on. On the other hand, conventional
quantum mechanics is recovered inside the light cone after an appropriate
relativistic delay. The practical implications of all this are yet to be seen.

\section{Superluminous propagation}\label{s:super}
The Feynman path integral formulation of quantum mechanics in one dimension
postulates that the propagation of a free particle is defined by the action
functional
 \beq
 S(x(t))=\int\limits_0^t\frac{m{\dot x}^2(t)}{2}\,dt\label{action}
 \eeq
as a convolution with the propagator
 \beq
 \exp\left\{\frac{i}{\hbar} S(x(t))\right\}.\label{Saction}
 \eeq
This means that the propagation of the time-sliced wave function at time
$t+\Delta t$ is given by  \cite{Keller}
 \beq
 \psi(x,t+\Delta t)=\sqrt{\frac{m}{{2\pi i\hbar\Delta t}}}\int\limits_{-\infty}^{\infty}
 \exp\left\{\frac{i m}{\hbar}\frac{(x-y)^2}{2\Delta
 t}\right\}\psi(y,t)\,dy.\label{psixt}
 \eeq
The limit $\Delta t\to0$ is found \cite{Feynman}, \cite{Keller} by changing the
variable of integration to $y=x-z\sqrt{\hbar\Delta t/m}$, which converts
(\ref{psixt}) to
  \begin{align}
 \psi(x,t+\Delta t)=&\,\sqrt{\frac{1}{{2\pi i}}}\int\limits_{-\infty}^{\infty}
 \exp\left\{\frac{iz^2}{2}\right\}\psi\left(x-z\sqrt{\frac{\hbar\Delta
 t}{m}},t\right)\,dz\label{psixtz0}\\
 =&\,\sqrt{\frac{1}{{2\pi i}}}\int\limits_{-\infty}^{\infty}
 \exp\left\{\frac{iz^2}{2}\right\}\times\nonumber\\
 &\,\left[\psi(x,t)-z\sqrt{\frac{\hbar\Delta
 t}{m}}\psi_x(x,t)+\frac12z^2\frac{\hbar\Delta
 t}{m}\psi_{xx}(x,t)+\cdots\right]\,dz\nonumber\\
 =&\,\psi(x,t)-\frac{i}{2}\frac{\hbar\Delta
 t}{m}\psi_{xx}(x,t)+O\left(\left(\frac{\hbar\Delta
 t}{m}\right)^2\right),\nonumber
 \end{align}
hence Schr\"odinger's equation follows.

If instead of the non-relativistic Lagrangian ${\cal L}(z)=z^2/2$ in
(\ref{action}), we use the unique relativistic Lagrangian of a free particle in
one dimension ${\cal L}(z)=-\sqrt{1-z^2}$ \cite{Miura}, it leads to the Lorentz
invariant action
 \beq
 S(x(t))=\int\limits_0^tmc^2{\cal L}\left(\frac{\dot x(t')}{c}\right)\,dt'.\label{actionf}
 \eeq
Now, the Lorentz invariant propagation is given by
 \beq
 &&\psi(x,t+\Delta t)={\cal N}^{-1}\times\label{psixtex0}\\
 && \int\limits_{-\infty}^{\infty} \exp\left\{i\left[1-\sqrt{1-\ds\frac{(x-y)^2}{c^2\Delta
t^2}}\right]\frac{mc^2\Delta t}{\hbar}\right\}H\left[c^{2}\Delta
t^{2}-(x-y)^{2}\right]\psi(y,t)\,dy,\nonumber
 \eeq
where the normalization constant ${\cal N}$ is given by
 \begin{align}
 {\cal N}=c\Delta t\int\limits_{-\infty}^{\infty}
 \exp\left\{\frac{i mc^2\Delta t}{\hbar}{\cal L}\left(\frac{x-y}{c\Delta
 t}\right)\right\}H\left[c^{2}\Delta t^{2}-(x-y)^{2}\right]\,dy.\label{N0}
 \end{align}
Note that modifying ${\cal L}(z)$ to ${\cal L}(z)=1-\sqrt{1-z^2}$ does not
change the integral equation (\ref{psixtex0}). Obviously, the wave function
$\psi(x,t)$ defined by (\ref{psixtex0}) vanishes on and outside the light cone.

Taking the first the limit $c\to\infty$ converts (\ref{psixtex0}) to
(\ref{psixt}), and then the limit $\Delta t\to0$, as in (\ref{psixtz0}),
recovers the Schr\"odinger equation in the entire space and the zero boundary
conditions on light cone disappear and so does the light cone. Evidently, the
wave function (\ref{psixt}) spreads to the entire line instantaneously, thus
rendering Schr\"odinger's equation superluminous. Taking the limits in the
reverse order does not lead to a wave equation.

The decay of the integral in (\ref{psixt}) for large $|x|$, similarly to the
Fourier transform, is determined by the regularity of $\psi(y,t)$. Thus jump
discontinuities in $\psi(y,t)$ lead to a decay of $\psi(x,t+\Delta t)$ as
$|x|^{-1}$, which prevents the existence of moments, energy, and other
artifacts \cite{PLA}.

\section{Subluminous propagation}

From the mathematical point of view, the approximation of the integral
(\ref{psixtex0}) by taking limits in the order described in Section
\ref{s:super} is a regular expansion of a singular approximation problem. The
singularity of this expansion is manifested in the loss of the boundary
conditions on the light cone. We propose here a distinguished limit that
removes the singularity  and the the ensuing artifacts. Specifically, we
consider a general Lorentz-invariant action functional (\ref{actionf}), where
the Lagrangian ${\cal L}(z)$ is an analytic function near the origin with
 \beq
 {\cal L}(0)=0,\quad  {\cal L}'(0)=0,\quad {\cal L}''(0)=1.\label{ffpfpp}
 \eeq
The assumption that the action (\ref{actionf}) is Lorentz-invariant implies
that the time-sliced wave function it defines cannot propagate outside the
light cone emanating from the support of the initial wave function. We
postulate, as above, the integral equation
 \beq
&& \psi(x,t+\Delta t)={\cal N}^{-1}\times\label{psixtc}\\
&&\int\limits_{-\infty}^{\infty}
 \exp\left\{\frac{i mc^2\Delta t}{\hbar}{\cal L}\left(\frac{x-y}{c\Delta
 t}\right)\right\}H\left[c^{2}\Delta t^{2}-(x-y)^{2}\right]
 \psi(y,t)\,dy,\nonumber
 \eeq
with the normalization factor ${\cal N}$ given by
 \beq
 {\cal N}&=&c\Delta t\int\limits_{-\infty}^{\infty}
 \exp\left\{\frac{i mc^2\Delta t}{\hbar}{\cal L}\left(\frac{x-y}{c\Delta
 t}\right)\right\}H\left[c^{2}\Delta t^{2}-(x-y)^{2}\right]\,dy\nonumber\\
 &=&c\Delta t\int\limits_{-1}^{1}
 \exp\left\{i{\cal L}(z)\xi\right\}dz,\label{N}
 \eeq
where
 \beq
 \xi=\ds\frac{mc^2\Delta t}{\hbar}.\label{xi}
 \eeq

\begin{figure}
\centering \resizebox{!}{7cm}{\includegraphics{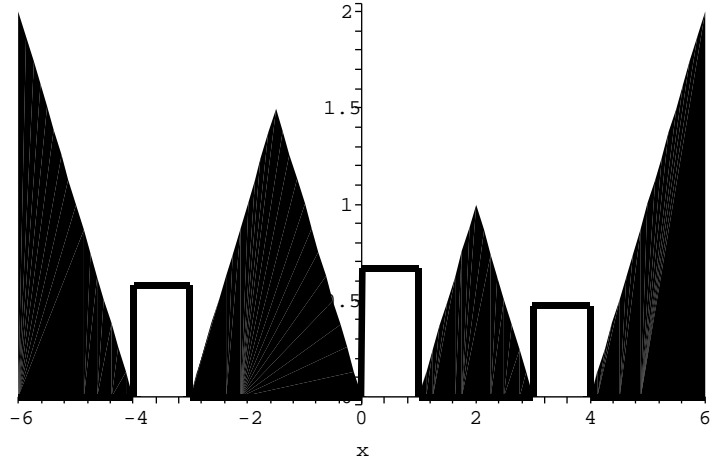}} \caption{\small The
thick line is the initial wave function, normalized in $L^2(\rR)$. The light
cone for $c=1$ and $t<2$ is the domain above the $x$-axis and above the black
region in the $(x,t)$ plane.} \label{f:Light-cone}
\end{figure}

To investigate the behavior of $\psi(x,t)$ on the  boundary of the
light cone emanating from the initial support, we rewrite
(\ref{psixtc}) in the explicit form
 \beq
\psi(x,t+\Delta t)={\cal N}^{-1}\int\limits_{x-c\Delta t}^{x+c\Delta t}
 \exp\left\{\frac{i mc^2\Delta t}{\hbar}{\cal L}\left(\frac{x-y}{c\Delta
 t}\right)\right\}
 \psi(y,t)\,dy.\label{psixplicit}
 \eeq
If $x_0$ is a point on the boundary of the initial support, we assume the light
cone emanating from it at time $\Delta t$ is $x=x_0-c\Delta t$. At this point
(\ref{psixplicit}) is
 \beq
  && \psi(x_0-c\Delta t,\Delta t)={\cal
N}^{-1}\times\nonumber\\
&&\nonumber\\ &&\int\limits_{x_0-2c\Delta t}^{x_0}
 \exp\left\{\frac{i mc^2\Delta t}{\hbar}{\cal L}\left(\frac{x_0-c\Delta t-y}{c\Delta
 t}\right)\right\}
 \psi(y,0)\,dy=0,\label{psixplicit1}
 \eeq
because the entire (open) interval of integration, $(x_0-2c\Delta t,x_0)$, is
outside the support of $\psi(y,0)$. Note that $\psi(x_0-c\Delta t,\Delta t)=0$,
even if $\psi(x_0,0)\neq0$. This boundary condition persists as long as the
line $x=x_0-ct$ is on the light cone (see Figure \ref{f:Light-cone}). Thus, if
$(a,b)$ is a finite interval on the $x$-axis outside the support of
$\psi(x,0)$, then the wave function $\psi(x,t)$ vanishes in the triangle in the
$(x,t)$ plane, whose vertices are the points $[a,0]$, $[b,0]$, and
$[(a+b)/2,(b-a)/2c]$. This implies, for example, that if in the two slits
experiment $(a,b)$ is the interval separating the two slits, there will be no
interference pattern between the slits before time $(b-a)/2c$. Furthermore, the
influence on the wave function of every finite interval in the support of the
initial wave function is confined to the light cone emanating from the interval
(see Figure \ref{f:Light-cone}).

It is easily seen that the solution $\psi(x,t)$ of (\ref{psixplicit}) is
Lorentz-invariant on the discrete time grid and it is supported on the light
cone emanating from the initial support.

\subsection{The distinguished limit}\label{ss:distinguished}

Equations (\ref{N}) and (\ref{xi}) suggest the distinguished limit of large $c$
and small $\Delta t$ such that $c\Delta t\to0$, but $c\sqrt{\Delta t}\to\infty$
(in appropriate dimensionless units). In more precise mathematical terminology
the meaning of this distinguished limit is the assumption that $\xi\to\infty$,
or more specifically,
 \beq
\frac{\hbar}{mc^2}\ll\Delta t\ll \frac{L^2m}{\hbar},\label{regiondt}
 \eeq
where $L$ is a characteristic length, e.g., the width of the initial support of
the wave function (the lower bound is due to the presence of $c$, and the
corresponding characteristic time and the upper is needed for the approximation
of a finite difference quotient by a derivative, see below).

For a point $(x,t)$ inside the light cone emanating from the initial support,
we change the variable of integration to $y=x-zc\Delta t$ and write
(\ref{psixtc}) as
 \beq
\psi(x,t+\Delta t) & =&{\cal N }^{-1}c\Delta t \times\label{integ}\\
&&\int\limits_{-1}^{1} \exp\left\{i{\cal L}(z)\frac{mc^2\Delta
t}{\hbar}\right\}\psi(x-c\Delta t(1+z),t)\,dz.\nonumber
 \eeq
If the light cone containing $(x,t)$ is the interval $x_0-ct<x<x_0+L+ct$, we
assume that $x_0-ct+\eps=x<x_0+L+ct$ for some positive $\eps$ independent of
$\Delta t$. Then integration in (\ref{integ}) extends over the interval
$-1<z<-1+\eps/c\Delta t$, which contains the entire interval $-1<z<1$ if
$c\Delta t $ is sufficiently small. Now we expand in Taylor's series to convert
(\ref{integ}) to
 \beq
\psi(x,t+\Delta t) & =&{\cal N }^{-1}c\Delta t
\int\limits_{-1}^{1} \exp\left\{i{\cal L}(z)\frac{mc^2\Delta t}{\hbar}\right\}\times\label{psixtcz}\\
&&\left[\psi(x,t)-zc\Delta t\psi_x(x,t)+\frac12z^2c^2\Delta
t^2\psi_{xx}(x,t)+\cdots\right]\,dz.\nonumber
 \eeq
The left hand side of the inequality (\ref{regiondt}) ensures that
$\xi\gg1$ (see (\ref{xi})), so the integrals in (\ref{psixtcz}) can
be evaluated by the stationary phase method \cite{Bender}. The right
hand side of (\ref{regiondt}) serves to ensure that $\psi(x,t+\Delta
t)-\psi(x,t)=\Delta t\psi_t(x,t)+o(\Delta t)$. Thus the expansion
(\ref{psixtcz}) with the assumptions (\ref{ffpfpp}),
(\ref{regiondt}) gives
\begin{align}
i\psi_t(x,t)=-\frac{\hbar}{2m}\psi_{xx}+O\left(\xi^{-1/2}\right),\label{schreq}
\end{align}
which in the (formal) limit $\xi\to\infty$ is Schr\"odinger's equation in every
interior point of the light cone.

Because the support of $\psi(x,t+\Delta t)$ cannot exceed the support of $\psi(x,t)$ by more than
$c\Delta t$ on either side, Schr\"odinger's equation (\ref{schreq}) with a finitely supported
initial wave function has to be solved inside the light cone emanating from the initial support
(see Figure \ref{f:Light-cone}) with vanishing boundary conditions. In particular, zero boundary
conditions have to be imposed on the triangles mentioned above. Thus, for example, if the initial
wave function vanishes on a finite number of finite intervals and outside the minimal finite
interval containing its support, the zero boundary conditions on the triangles are given only for a
finite time, but the zero boundary condition on the light cone emanating from the minimal interval
containing the support are given for all times. The Schr\"odinger equation has to be solved
piecemeal outside the triangles for finite time intervals and then in the entire light cone
emanating from the above mentioned minimal interval. Obviously, the initial conditions at times
corresponding to every apex of a triangle, as mentioned above, is the wave function obtained from
solving the Schr\"odinger equation in the light cones emanating from the support up to that time.
Thus, after the time corresponding to the highest triangle the Schr\"odinger equation has to be
solved in the light cone emanating from the minimal interval, with the initial value of the wave
function constructed up to that time. An important result of this structure is that at all times
the wave function is analytic inside its (finite) support and has finite moments at all times.

\begin{figure}
\centering \resizebox{!}{6cm}{\includegraphics{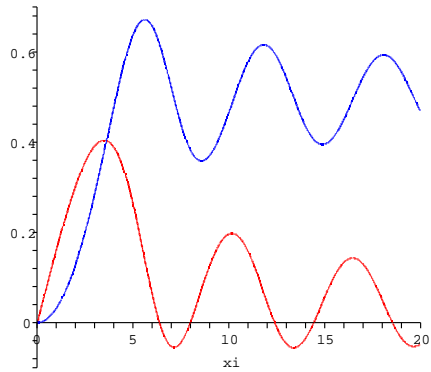}} \caption{\small The
lower curve is the real part and the upper is the imaginary part of the
coefficient of $\frac{\hbar}{m}\psi_{xx}(x,t)$ in (\ref{psixtz}) for
$0<\xi<20$} \label{f:Variance20}
\end{figure}

\begin{figure}
\centering \resizebox{!}{6cm}{\includegraphics{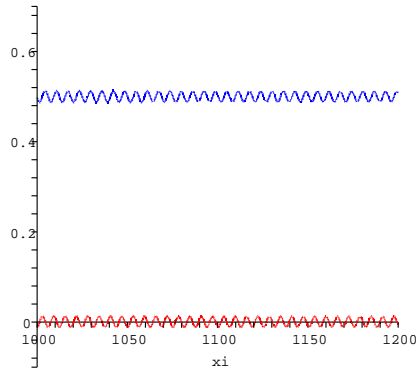}} \caption{\small
The lower curve is the real part, oscillating around its limit $0$ and upper
curve is the imaginary part, oscillating around the limit value of the
coefficient $\frac12$ of $\frac{\hbar}{m}\psi_{xx}(x,t)$ in (\ref{psixtz}) for
$1000<\xi<1200$.} \label{f:Variance1200}
\end{figure}

We apply the above procedure to the case discussed in Section \ref{s:super}
${\cal L}(z)=1-\sqrt{1-z^2}$. In this case all integrals can be expressed
explicitly in terms of Bessel and Struve functions \cite{Stegun} without
invoking the saddle point expansion. The Lorentz invariant propagation is now
given by (\ref{psixtex0}), so the change the variable of integration to
$y=x-zc\Delta t$ and expansion in Taylor's series converts (\ref{psixtex0}) to
 \beq
\psi(x,t+\Delta t) & =&{\cal N}^{-1}c\Delta t
\int\limits_{-1}^{1} \exp\left\{-i\left[\sqrt{1-z^2} -1\right]\frac{mc^2\Delta t}{\hbar}\right\}\times\label{psixtz}\\
&&\left[\psi(x,t)-zc\Delta t\psi_x(x,t)+\frac12z^2c^2\Delta
t^2\psi_{xx}(x,t)+\cdots\right]\,dz.\nonumber
 \eeq
The coefficient of $\ds\frac{\hbar}{m}\psi_{xx}(x,t)$ in (\ref{psixtz}) is
shown in Figures \ref{f:Variance20} and \ref{f:Variance1200}.
In the limit $\Delta t\to0$, that is, $\xi\to0$, all coefficients in the
expansion (\ref{psixtz}) (after dividing by $\Delta t$) converge to $0$, giving
$\psi_t(x,t)=0$ so the path integral converges to the initial wave function.
However, the limit $c\Delta t\to0$ and $\xi\to\infty$ in the expansion
(\ref{psixtz}) yields (\ref{schreq}). It is evident that the path integral
converges to the solution of the subluminous Schr\"odinger equation as $\Delta
t\to0$, except for a narrow range of $\Delta t$, where the restriction
(\ref{regiondt}) is violated.

\section{Quantization}
In the presence of a potential $\Phi(x)$ the action (\ref{actionf}) is replaced
with
 \beq
 S(x(t))=\int\limits_0^t\left[mc^2{\cal L}\left(\frac{\dot x(t')}{c}\right)-\Phi(x(t'))\right]\,dt',\label{actionV}
 \eeq
and the exponent in (\ref{psixtc}) becomes
 \beq
\frac{i}{\hbar} \left\{mc^2{\cal L}\left(\frac{x-y}{c\Delta
 t}\right)-\Phi(y)\right\}\Delta t.
 \eeq
The Schr\"odinger approximation is now
 \beq
 i\hbar\psi_t(x,t)=-\frac{\hbar^2}{2m}\psi_{xx}(x,t)+\Phi(x)\psi(x,t)+O\left(\xi^{-1/2}\right)\label{SEQ}
 \eeq
within the appropriate light cones confines.
 Quantized energy levels appear when
the potential has a proper local minimum. To see how discrete energies emerge
in the subluminous Schr\"odinger equation, we consider the elementary case of
the harmonic oscillator with potential $\Phi(x)=\frac12m\omega^2x^2$ and assume
that the support of $\psi(x,0)=\varphi(x)$ is the interval $(-x_0,x_0)$ and
that the boundary conditions are $\psi(\pm (x_0+ct),t)=0$. First, we
nondimensionalize  equation (\ref{SEQ}) by setting
 \beq
 x=y\sqrt{\frac{\hbar}{m\omega}},\quad t=\frac{\tau}{\omega},\quad
 \psi(x,t)=U(y,\tau),\quad \varepsilon=\frac1c\sqrt{\frac{\omega\hbar}{m}},
 \eeq
which converts the initial and boundary value problem for Schr\"odinger's
equation (\ref{SEQ}) into
 \beq
2iU_\tau(y,\tau)&=& -U_{yy}(y,\tau)+y^2U(y,\tau)\hspace{0.5em}\mbox{for}\ |y|<y_0+\frac\tau\eps\label{UEQ}\\
&&\nonumber\\
U(y,0)&=&\varphi\left(y\sqrt{\frac{\hbar}{m\omega}}\right),\hspace{0.5em}
U\left(\pm\left(y_0+\frac\tau\eps\right),\tau\right)=0.\label{IBC}
 \eeq
Introducing the fast and slow time scales $\sigma=\tau/\eps$ and $\tau$,
respectively, and setting $U(y,\tau)=V(y,\sigma,\tau)$, we rewrite (\ref{UEQ})
as
 \beq
 2i(V_\tau+\eps^{-1}V_\sigma)=V_{yy}+y^2V
 \eeq
and assume the regular outer expansion \cite{Bender}
 \beq
 V=V^0+\eps V^1+\cdots.\label{outer}
 \eeq
The resulting hierarchy of equations
 \beq
 V^0_\sigma=0,\quad 2iV^1_\sigma=2iV^0_\tau-V^0_{yy}+y^2V^0,
 \dots\label{Hierarchy}
 \eeq

implies that $V^0$ is independent of $\eps$ and $\sigma$ and thus
cannot satisfy the boundary condition on the light cone
$y=\pm(y_0+\sigma)$. The solvability condition for
(\ref{Hierarchy}), obtained from averaging with respect to the fast
variable $\sigma$, is the Schr\"odinger's equation
\begin{align}
2iV^0_\tau-V^0_{yy}+y^2V^0=0,
\end{align}
which has to be solved on the entire line with the initial condition
(\ref{IBC}). The first approximation $V^0$ is thus merely the scaled
classical quantum harmonic oscillator
 \beq
V^0(y,\tau)= \psi^0(x,t)=\sum_{n=0}^\infty
a_ne^{-iE_nt/\hbar}\psi_n(x),
 \eeq
where $E_n$ are the energies, $\psi_n(x)$ are the (real valued)
eigenfunctions of the harmonic oscillator on the entire line and
$a_n$ are the coefficients in the expansion of the initial wave
function $V^0(y,0)=\varphi(x)$ in the eigenfunctions $\psi_n(x)$.

Higher order terms in the outer expansion (\ref{outer}) cannot
satisfy the boundary conditions (\ref{IBC}), so a boundary layer
correction \cite{Bender} is needed at $y=\pm(y_0+\tau/\eps)$.
 We thus introduce the
boundary layer variable and function near the right boundary
 \beq
 \eta=\frac{y_0+\ds\frac\tau\eps-y}{\eps},\quad U(y,\tau)=W(\eta,\tau)
 \eeq
Schr\"odinger's equation (\ref{SEQ}) gives
 \beq
 2i\eps^2W_\tau+2iW_\eta=-W_{\eta\eta}+\left(\eps y_0+\tau-\eps^2\eta\right)^2W.
 \eeq
Expanding $W=W^0+\eps W^1+\cdots$, we obtain the boundary layer
equation
 \beq
2iW^0_\eta&=&-W^0_{\eta\eta}+\tau^2W^0
 \eeq
  and the boundary and matching conditions
 \beq
W^0(0,\tau)&=&-V^0\left(y_0+\frac\tau\eps,\tau\right),\quad
W^0(\infty,\tau)=0.
 \eeq
The solution is given by
 \beq
 W^0(\eta,\tau)=-V^0\left(y_0+\frac\tau\eps,\tau\right)\exp\left\{-\eta\left[i+\sqrt{\tau^2-1}\,\,\right]\right\}.
 \eeq
With a similar construction at the left end of the interval, we get the uniform
expansion for $\eps\ll1$
 \beq
U(y,\tau)&\sim&H\left[(y-y_0)^2-\frac{\tau^2}{\eps^2}\right]{\Bigg[}V^0(y,\tau)\label{Uyt}\\
&&-V^0\left(y_0+\frac\tau\eps,\tau\right)
\exp\left\{-\left[i+\sqrt{\tau^2-1}\,\,\right]\frac{\tau+\eps(y_0-y)}{\eps^2}\right\}\nonumber\\
&&-V^0\left(-y_0-\frac\tau\eps,\tau\right)\exp\left\{-\left[i+\sqrt{\tau^2-1}\,\right]
\frac{\tau+\eps(y_0+y)}{\eps^2}\right\}{\Bigg]}.\nonumber
 \eeq

Equation (\ref{Uyt}) indicates that, for large $\tau/\eps$, the
expansion of the wave function in the eigenmodes of Schr\"odinger's
equation on the line is recovered inside the light cone, away from
the boundary. To regain a single mode, the initial wave function has
to be of the boundary layer form (\ref{Uyt}) at time $\tau=0$ and
$V^0(y,0)$ has to be the an eigenfunction of Schr\"odinger's
equation on the entire line.

Obviously, energy is preserved, because boundary conditions and integration by
parts of the energy integral give
 \beqq
 \frac{dE(t)}{dt}=-ic\psi\bar\psi_t(x_0+ct)+ic\psi\bar\psi_t(-x_0-ct)-
 i\int\limits_{-x_0-ct}^{x_0+ct}[\psi_t\bar\psi_t+\psi\bar\psi_{tt}]\,dx=0.
 \eeqq
The energy at time $t$ is up to errors of order of magnitude of the
contribution of the boundary layer
 \begin{align}
 E(t)\sim&\int\limits_{-x_0-ct}^{x_0+ct}\sum_{m,n=0}^\infty
 \left[-a_m\bar a_ni\hbar\frac{\p}{\p t}e^{-iE_mt/\hbar}e^{iE_nt/\hbar}\psi_m(x)\bar\psi_n(x)\right]\,dx\nonumber\\
=&\sum_{m,n=0}^\infty a_m\bar a_n
E_me^{i(E_n-E_m)t/\hbar}\int\limits_{-x_0-ct}^{x_0+ct}\psi_m(x)\bar\psi_n(x)\,dx\nonumber\\
&\to\sum_{m=0}^\infty|a_m|^2 E_m\quad\mbox{as}\ ct\to\infty.\label{Et}
 \end{align}

\section{Discussion}

The standard non-relativistic Schr\"odinger model of quantum mechanics consists
of the initial value problem for Schr\"odinger's equation in the entire space.
In the conventional interpretation of this model any boundary conditions
imposed on the wave function can be represented in terms of appropriate
potentials in Schr\"odinger's equation. In this paper we adopt an alternative
model, which consists of a Lorentz-invariant Feynman path integral that
perforce vanishes on and outside the light cone emanating from the support of
the initial wave function. This boundary condition does not require any
potentials, rather, it is an inseparable part of this model of quantum
mechanics. As explained in this paper, the path integral converges to a limit
as the time slice $\Delta t\to0$, except in a narrow range where the
restriction (\ref{regiondt}) is violated. The distinguished limit $c\Delta
t\to0$ and $\xi\to\infty$ of the path integral is the solution of the
subluminous Schr\"odinger equation (the solution that satisfies zero boundary
conditions on and outside the original light cone of the path integral). We can
view, therefore, the subluminous Schr\"odinger equation as an intermediate
model between non-relativistic and relativistic quantum mechanics. The results
of the standard non-relativistic quantum theory are recovered in the
intermediate model, albeit after a relativistic delay.

To clear the issue of how the results of this paper relate to existing
theories, we point that
\begin{enumerate}
\item This paper mainly addresses issues related to the original (non-relativistic)  Schr\"odinger equation.
\item Ours is not a relativistic quantum theory, but rather a conventional one
that upholds the basic tenets of physics: all propagation must be confined
within a relevant light-cone.
\item Why should one care about such matters as light-cones in the non-relativistic regime?
The quick answer is that neglecting the speed of light and the
resulting boundary conditions on light-cones is, mathematically
speaking, a singular perturbation, which results in unphysical
artifacts such as the disappearance of moments and energy when the
initial wave function is discontinuous (e.g., after a collapse due
to a measurement) or the instantaneous spread of information into
the entire space. These artifacts merely reflect the mathematical
shortcomings ingrained in the original Schr\"odinger formulation.
\item Recall that similar difficulties occur in diverse physical theories.
For instance, air viscosity may be practically negligible, but if it is
neglected altogether, gas dynamics predicts no drag force on airplanes wings.
For no matter how irrelevant it is elsewhere, close to a moving body there is a
boundary layer, where viscous effects are dominant. Thus, while viscosity
manifests itself locally, its effect is global!
\item A closer kin to our problem  is the heat equation and its unphysical artifact of instantaneous
spreading of initially localized perturbation to the entire line. This artifact
can be attributed to the disregard of the acoustic speed limitation in the
derivation of the diffusion equation. If the acoustic bound is enforced on the
process, finite propagation fronts emerge and the unphysical artifact of
instantaneous propagation disappears \cite{PR1}.
\item Perhaps the most succinct summary of our discussion would be
to say that it is not enough that we know the limitation of our model; the
model itself should contain its own limitations, as is the case at hand:  the
Schr\"odinger equation with boundary conditions contains the limitation on
propagation.
\item The non-relativistic propagator of Schr\"odinger's equation has some of
the properties of the Fourier transform: the decay at infinity of the wave
function reflects the smoothness of its initial value. Thus a jump
discontinuity is converted by the propagator into a decay of the wave function
at infinity as $O\left(|x|^{-1}\right)$, which although square integrable, has
no moments \cite{PLA}. In contrast, none of this anomaly exists for
Schr\"odinger's equation with zero boundary conditions on and outside the light
cone.
\end{enumerate}

To conclude, while confining the  Schr\"odinger equation within a relevant
light cone may only mildly extend the scope of its applicability, which is yet
to be seen, it definitely relieves it of annoying paradoxes and artifacts.
Given the central position of Schr\"odinger's equation in modern physics, our
extension should be expected to have implications beyond mere scientific
esthetics. The basic predicament of relativity manifests itself in our theory
only in crucial 'junctions' of space-time: the theory limits the domain of
propagation, endows the wave function with a definite front, thus eliminating
the unlimited precursors and causing a time delay for the conventional quantum
effects to re-emerge.\\

\noindent{\bf Acknowledgments}. We thank A. Marchewka for useful discussions.
The work of the first author was supported by the ISF grant no.801/07.

\end{document}